05/11/25

# PROTEIN AGGREGATION IN HUNTINGTON'S DISEASE

Guylaine Hoffner and Philippe Djian*

*To whom correspondance should be addressed
E-mail: philippe.djian@biomedicale.univ-paris5.fr
Fax: 01 42 86 20 61
Tel: 01 42 86 22 72

Key words: Aggregates, inclusions, Huntington's disease, polyglutamine.

UPR 2228 – CNRS
Institut Interdisciplinaire des
Sciences du Vivant des Saints-Pères
Université René Descartes
45, rue des Saints-Pères
75270 Paris Cedex 06
France

**Abstract**

The presence of an expanded polyglutamine produces a toxic gain of function in huntingtin. Protein aggregation resulting from this gain of function is likely to be the cause of neuronal death. Two main mechanisms of aggregation have been proposed: hydrogen bonding by polar-zipper formation and covalent bonding by transglutaminase-catalyzed cross-linking. In cell culture models of Huntington's disease, aggregates are mostly stabilized by hydrogen bonds, but covalent bonds are also likely to occur. Nothing is known about the nature of the bonds that stabilize the aggregates in the brain of patients with Huntington's disease. It seems that the nature of the bond stabilizing the aggregates is one of the most important questions, as the answer would condition the therapeutic approach to Huntington's disease.

## Introduction

The cause of Huntington's disease is the expansion of a polyglutamine sequence close to the N-terminus of huntingtin, a protein of unknown function [1]. Huntington's disease is one of nine diseases caused by polyglutamine expansion in nine unrelated proteins. Diseases of polyglutamine expansion are characterized by the presence of microscopic inclusions in neurons of the central nervous system (figure 1). These inclusions, whose average size is about 5-7 μm, are formed by aggregates of the protein with the expanded polyglutamine sequence [2, 3]. A number of questions have been raised by the discovery of the inclusions : what is the mechanism of aggregation of proteins containing expanded polyglutamine ? Are inclusions the cause of neuronal death ? What is the composition of the inclusions ? Why are inclusions often nuclear or perinuclear ?

**Mechanism of aggregation of proteins bearing expanded polyglutamine.**

In 1993, Howard Green hypothesized that, as a result of excessive polyglutamine length, huntingtin could become aggregated by transglutaminase [4]. Transglutaminases catalyze the formation of isopeptide cross-links between the side chains of glutaminyl and lysyl residues of adjacent proteins. In 1994, Max Perutz proposed that polyglutamine sequences could form multimers stabilized by hydrogen-bonded polar zippers between the main chain and side chain amides [5]. These two hypotheses may not be mutually exclusive.

***1. Polar zippers***

The first evidence for polar zipper formation by polyglutamine sequences came from the discovery that incorporation of $Q_{10}$ into a small protein resulted in the formation of oligomers [6]. It was later shown that transgenic mice expressing the first exon of human huntingtin, which encoded about 130 glutamine residues flanked by 17 N-terminal and 52 C-terminal residues developped neurological disease and neuronal inclusions [7]. The bacterially-produced peptide encoded by exon 1 formed aggregates in vitro when its polyglutamine sequence exceeded 51 residues, but not when the polyQ was 20 or 30 residues [8]. Since purified peptide was used, aggregation must have resulted from multimerization by polar zipper formation. The authors concluded that the inclusions found in exon 1 transgenic mice were similarly stabilized by hydrogen-bonded polar zippers, but there was no direct analysis of the intermolecular bonds found in the inclusions of mice.

Hazeki et al. [9] introduced the N-terminal fragment of huntingtin bearing $Q_{77}$ into cultured cells by transfection, and observed the formation of inclusions. Incubation of the cell extract in the presence of 100% formic acid reduced most of the aggregates to monomer (65 kD). However a band corresponding to the size of a dimer (130 kD) was also released. It can be concluded that most of the aggregates were stabilized by hydrogen bonds. The 130 kD band might have resulted from cross-linking of the poly Q-containing peptide, either to itself or to another peptide. It is therefore clear from these experiments that in cultured cells most of the aggregation resulted from polar-zipper formation, but that a participation of transglutaminase catalyzed cross-linking is likely.

No direct evidence has yet been provided as to the existence of polar zippers in the aggregates or inclusions present in the brain of patients. Hydrogen-bonded β pleated sheets are known to exhibit green birefringence under polarized light after Congo red staining. Aggregates formed by a pure peptide containing an expanded polyQ sequence also display such a green birefringence [8]. The fact that some of the inclusions found in the brain of patients with Huntington's disease were birefringent after staining with Congo red has been taken as evidence that these inclusions contained β pleated sheets stabilized by hydrogen bonds [10, 11]. However Karpuj et al. were unable to stain the inclusions of patients with Congo red [12] and McGowan et al. have reported that the inclusions present in the exon 1 transgenic mice cannot be stained with Congo red [11].

### *2. Transglutaminase-catalyzed cross-linking*

Transglutaminase is widely distributed in the brain [13, 14] and it can be activated, since ε-(γ-glutamyl) lysine cross-links are found in normal brain [15]. Three different transglutaminases (numbered 1, 2 and 3) have been found in neurons [14]. Transglutaminase is thought to be increased in the brain of patients with Huntington's disease [16]. The demonstration of transglutaminase in the nuclei of neurons has added plausibility to the fact that nuclear inclusions could result from enzymatic cross-linking [11].

There is ample evidence that polyglutamine is a substrate of transglutaminase in vitro. Model peptides containing $Q_5$-$Q_{18}$ have been shown to be much better substrates than those containing $Q_1$-$Q_3$ [17]. Synthetic peptides composed of pure polyQ with no flanking sequence accept spermine in proportion to the length of the polyQ sequence [18]. $Q_{10}$ and $Q_{62}$ fused to glutathione S-transferase are better substrates than dimethylcasein [19]. Intact huntingtin is a substrate of transglutaminase in vitro and the rate of reaction increases with the length of the polyQ sequence [20]. Karpuj et al. [11] found that in vitro translated exon 1 huntingtin was soluble even when it contained $Q_{67}$, and that transglutaminase cross-linked the exon 1 protein into high molecular weight polymers. More polymers formed with $Q_{67}$ than with $Q_{23}$. This was in contrast to the finding of Scherzinger et al. [8] that a similar peptide containing $Q_{51}$ aggregated spontaneously in the absence of transglutaminase.

There is yet no direct evidence that the inclusions found in brain of patients contain cross-links. However, in a recent report, the levels of ε-(γ-glutamyl) lysine in spinal fluid was found to be increased by three fold in patients with Huntington's disease [21]. The free isopeptide represents the ultimate product of proteolytic digestion of cross-linked proteins. Its abundance in spinal fluid is therefore likely to reflect the abundance of cross-linked peptides in brain.

The role of transglutaminase in Huntington's disease has been recently challenged. Human neuroblastoma cells, stably transformed with a transglutaminase-2 antisense construct, formed aggregates when they were transiently transfected with an expression vector encoding an N-terminal fragment of huntingtin with $Q_{82}$. These cells contained no detectable transglutaminase-2, but it was not clear whether they contained transglutaminases-1 and -3 [22].

**Are inclusions the cause of neuronal death ?**

It seems likely that aggregation of huntingtin is the cause of neuronal death in Huntington's disease since there is a direct correlation between frequency of inclusions in neurons and severity of the disease [23, 24] and since in transgenic mice, inclusions precede the disease [2]. However, some studies have challenged the role of huntingtin-containing inclusions in neuronal death.

Transfection of immature striatal neurons prepared from rat embryos with an expression vector encoding the first 171 residues of huntingtin with $Q_{68}$ resulted in ubiquitinated intranuclear inclusions and cell death. Inhibition of ubiquitination prevented the formation of inclusions but did not affect the rate of cell death. The authors conclude that "inclusions is not likely to play a positive role in mediating …. cell death" [25]. The conclusion of the authors might be right. However, microscopic inclusions represent the final stage of protein aggregation. Preventing ubiquitination of huntingtin might delay formation of inclusions without affecting aggregation. Submicroscopic aggregation of huntingtin such as that described by Gutekunst et al. [26] would have remained undetected in the experiments reported by Saudou et al.. It is also worth noting that polyglutamine sequences do not need to be ubiquitinated to form insoluble aggregates in vitro, and that a significant fraction of the inclusions found in neurons of patients are not ubiquitinated [3].

Transgenic *Drosophila* expressing in their brain a peptide with $Q_{78}$ showed progressive neurodegeneration and early lethality. Coexpression of HSP 70 decreased the neurodegeneration and delayed lethality, without affecting the onset, size or number of neuronal inclusions. The authors concluded that either the inclusions were not toxic or that their toxic effect was diminished by overexpression of HSP 70 [27].

**The protein composition of the inclusions**.

In the inclusions, the huntingtin containing an expanded polyglutamine appears to be associated with a variety of accessory proteins. It has been suggested that sequestration in the inclusions of proteins important for cell viability could have a role in the pathogenesis of the disease by decreasing the cytoplasmic or nuclear concentration of these proteins. Two main categories of accessory proteins have been found in inclusions : transcription factors and components of the ubiquitin-proteasome system pathway.

Suhr et al. [28] have introduced by transient transfection into 293 cells an inducible expression vector encoding the exon 1 fragment of huntingtin with $Q_{96}$ and a nuclear localization signal. Cytoplasmic and nuclear inclusions were purified 5 days after transfection, using a CsCl gradient, and incubated in gel loading buffer containing SDS and β-mercaptoethanol. The proteins released during the incubation were resolved by electrophoresis on denaturing polyacrylamide gels. Protein bands were then analyzed by Western blotting and mass spectrometry. Among the proteins found were ubiquitin, p53, mdm2, HSP70, TBP, actin, NFL and proteins of the nuclear pore complex. All of these proteins except actin and NFL were confirmed to be associated with the inclusions by immunocytochemistry. In these experiments, inclusions were not likely to have been solubilized since the $Q_{96}$ monomer remained aggregated at the top of the gel. It is therefore possible that a large fraction of the proteins present in the inclusions escaped detection. TBP has also been detected by immunohistochemistry in inclusions of patients with Huntington's disease [10]. Other transcription factors identified in the inclusions of cultured cells, transgenic mice or brains of patients are CBP [29, 30, 31] and mSin3a [32].

Heat shock proteins (HSP70 and 40), the 20S proteasome and ubiquitin colocalize with inclusions in cells expressing the exon-1 fragment of huntingtin. Heat shock and proteasome inhibition increase the number of inclusions [33]. Conversely, the presence of an aggregated polyQ peptide inhibits the ubiquitin-proteasome system. It is suggested that the presence of an excessively long polyQ sequence confers an abnormal conformation to the protein bearing the long polyQ sequence and thereby promotes its ubiquitination. The proteasomes would then become saturated with ubiquitinated aggregates that they cannot degrade and would therefore be unable to process their normal substrates [34]. However the fact that the inclusions disappear when induction of expression is discontinued in a conditionnal transgenic mouse model of Huntington's disease shows that the inclusions can be degraded [35].

It still remains to be determined whether the ubiquitin present in the inclusions is attached to the peptide bearing the expanded polyQ or to other proteins that are targeted for degradation and tend to coaggregate with the polyglutamine sequence. The recruitement of components of the ubiquitin-proteasome degradation pathway is not specific to the inclusions found in diseases of polyglutamine expansion since it is also observed in the neuronal Lewy bodies characteristic of Parkinson's disease, in neuronal Pick bodies and in the Mallory bodies present in hepatocytes suffering from ethanol toxicity.

**Are inclusions formed by N-terminal fragments of huntingtin ?**

N-terminal fragments of huntingtin bearing an expanded polyglutamine are more cell lethal and more readily form inclusions than the intact protein when they are expressed in cells or in transgenic animals [36, 37]. Inclusions present in the brain of patients with Huntington's disease are stained by antibodies directed against the residues of huntingtin located in the vicinity of the expanded polyQ, but not by those raised against more C-terminal parts of huntingtin which include over 3000 residues [3, 24]. This has been taken as evidence that the inclusions contain N-terminal fragments of expanded huntingtin and no intact protein. In recent experiments, Dyer et al [38] have reported that in brains of patients with Huntington's disease, there are aggregates insoluble in SDS and urea, which contain intact expanded huntingtin. This would seem in contradiction with the idea that aggregates are formed by N-terminal fragments of expanded huntingtin. It was not clear however, whether the insoluble material studied by Dyer et al. corresponded to the inclusions.

**Why are inclusions often nuclear or perinuclear ?**

Huntingtin is found in the cytosol where it combines with microtubules [39, 40]. Microtubules converge around the nucleus. The microtubule-binding property of huntingtin therefore explains why huntingtin is more abundant around the nucleus than in the rest of the cytosol. From its perinuclear location, an N-terminal fragment of huntingtin would readily gain access to the nucleus. The glutamine expansion would promote its insolubility and produce an inclusion [40]. N-terminal fragments of huntingtin have recently been found in the nuclei of fibroblasts prepared from patients with Huntington's disease [41]. It remains to be explained why inclusions are mostly nuclear in patients with juvenile Huntington's disease, whereas they are mostly cytoplasmic in those with the adult form of the disease [3].

Other studies have argued that inclusions formed by polyglutamine sequences are analogous to aggresomes [42] whose perinuclear location results from dynein-dependent retrograde transport of microaggregates on microtubules [43]. However disruption of microtubules in a cell culture model of Huntington's disease did not affect formation of the inclusions [44].

**Concluding remarks**

Although it seems probable that protein aggregation is the cause of cell lethality in Huntington's disease, a number of questions remain unanswered.

- How can huntingtin, which is a cytoplasmic protein, form aggregates in the nucleus, when its polyQ is in excess of about 60 residues, as patients with juvenile Huntington's disease.
- Are aggregates composed of N-terminal fragments of huntingtin or of the intact protein.
- What kind of intermolecular bond stabilizes the aggregates.

The answer to the last two questions will certainly require the purification of the aggregates, their solubilization and their biochemical analysis by very sensitive methods, such as mass spectrometry.

**Legend to Fig.1 Inclusions in brain of patients with Huntington disease.** Frozen sections were stained with an antibody directed against the N-terminus of huntingtin (green) and the nuclei were counterstained with Hoechst 33258. (A) In a juvenile case, staining of Brodmann area 7 of the cerebral cortex with the anti-huntingtin antibody reveals the presence of numerous nuclear inclusions. (B) Double staining of a protein specific to neuronal nuclei, NeuN (red) shows that virtually all inclusions reside in neurons. (C) No inclusions are present in the cerebellar cortex of the same patient. (D) In a representative section of Brodmann area 9 of an adult case, only one large cytoplasmic inclusion is visible. Bar graphs: A, C and D (10 μm), B (20 μm).